\documentclass[referee]{aa} % for a referee version
\usepackage{graphicx}
\usepackage{txfonts}
\begin{document}
   \title{Optimal slit orientation for long multi-object spectroscopic exposures}

   %\subtitle{I. Overviewing the $\kappa$-mechanism}

   \author{G. P. Szokoly
          \inst{1}
          }

   \offprints{G. P. Szokoly}

   \institute{Max-Planck-Institut f\"ur extraterrestrische Physik,
      	      Giessenbachstra{\ss}e, D-85748 Garching, Germany \\
              \email{szgyula@mpe.mpg.de}
             }

   \date{Received December 17, 2004; accepted June 20, 2005}

   \abstract{Historically, long-slit spectroscopic observations
were carried out using the parallactic angle for the slit orientation
if slit loss was an important consideration (either to maximize
the signal-to-noise or to do spectrophotometry). This 
requires periodic realignment of the slit position angle as the parallactic 
angle changes. This is not possible for multi-slit
observations where one slit position angle must be chosen for
the entire exposure. Common wisdom suggests using the parallactic
angle at the meridian (HA=0). In this paper, I examine what
the best strategy is for long, multi-slit exposures. I find that in
extreme cases (very long exposure time) the best choice is
to orient the slit \emph{perpendicular} to the parallactic
angle at the meridian. There are two effects to consider: the increasing
dispersion with increasing airmass
and the changing angle between the parallactic angle and the slit.
In the case of \emph{traditional} slit orientation, the two
effects amplify each other, thus rendering a significant
fraction of the observation useless. Using the perpendicular orientation,
the two processes work against each other, thus most of the
observation remains useful.
I will use, as an example, our 8 hour Lockman Hole observations
using the Keck telescope, but generic methods are given to
evaluate a particular observation.
I also make the tools available to the community.
   \keywords{atmospheric refraction, spectroscopy
               }
   }

   \maketitle
%
%________________________________________________________________

\section{Introduction\label{intro}}

The introduction of multi-object spectrographs (MOS) in optical
spectroscopy is quite often thought of as just a set of
traditional long-slit spectrographs (LS) and some consequences
are overlooked. The main difference between MOS and LS
spectroscopy is the constraints imposed by the geometry of the
instrument. While in LS spectroscopy, the slit orientation can
be chosen arbitrarily, this is no longer the case in MOS
spectroscopy: slit position angles can not be changed during a
set of observations.

As observers tend to concentrate on very faint objects (typical R-band
magnitude of 23-24 with 8-10 meter class telescopes), the integration
times are getting longer and longer. Sometimes a whole night (8-10
hours or more) is spent on a single mask. These observations
require a rethinking of the optimal observation strategy.

Since many MOS instruments lack an atmospheric dispersion
corrector (e..g. VIMOS on VLT, DEIMOS on Keck), atmospheric
dispersion is a serious problem.  Long observations (many hours)
span a large range of zenith distance.  Thus, a differential
refraction of a few arc seconds is quite common.

As most observations require high signal to noise or good spectral
resolution, using sufficiently wide slits to compensate for
this effect is not acceptable. Very wide slits have a devastating
effect on background limited exposures (since the sky background
grows linearly with slit width, while the object signal grows
much slower) and a wide slit also blurs the spectra.

For short exposures, one can observe `close to the parallactic angle',
i.e. align the slit with the atmospheric dispersion direction. This way
the photons from the object enter the slit, the dispersion only
introduces an additional tilt in the resulting spectra, which is
easy to correct for for most applications. If the goal of the observation
is to extract spatial information, too, then extra care is
required to correct for this effect.

In the case of longer
observations, the direction of dispersion projected on the sky,
i.e. the parallactic angle, varies in time. For single object
observations (i.e. long slit spectroscopy), one can compensate
by periodically realigning the slit. For MOS exposures, this is
not possible. For masks, a single slit orientation \emph{must} be
chosen for the whole exposure. 

In this paper, I examine how the effective slit loss can be estimated
and I demonstrate the effect using our sample observation of the
Lockman Hole using the Keck telescope. I also describe how to
use our Web-based service to find the best strategy for a particular
observation. 

I will start with
the current best determination of the atmospheric refraction. I also
work out a simplified formula that is sufficient in many applications.
I simulate different observational strategies and show the
effect of atmospheric refraction on the efficiency of observations.

\section{Atmospheric dispersion and large telescopes}

The importance of atmospheric refraction in spectroscopic observations
was emphasized by Filippenko (\cite{filippenko1982}). The paper
discusses the optimal strategy (and the effect of non optimal
strategies) for \emph{short} long-slit spectroscopic observations.
Even though the paper uses a formula to calculate the index of refraction
that became obsolete, it is still the strategy to be followed for
\emph{short} integrations. The refined formula to calculate the
refraction introduces only negligible changes. On the other hand, the
paper does not discuss the optimal strategy for \emph{long} integrations.

Cohen and Cromer (\cite{cohen1988}) calculates the magnitude of differential
refraction for the Keck and the Norris spectrographs for realistic
observing scenarios. The paper determines the limits beyond which
the atmospheric dispersion degrades the data, but does not discuss how
to optimize observations that go beyond these limits.

Donnelly et. al (\cite{donnelly1989}) discusses optimal observational
strategies for \emph{fiber} spectrographs for \emph{long} exposures. 
Unfortunately these results can not be directly applied to slit
spectrographs: In many (but not all) projects, the slit orientation
can be chosen arbitrarily, thus, there is an extra degree of freedom to
minimize the effect of atmospheric refraction. This is not possible
for fiber spectrographs, thus, this is not discussed in this paper.

\begin{table}
\caption{Current and future 8m class telescopes and MOS instruments.}
\label{teltable}      % is used to refer this table in the text
\centering                          % used for centering table
\begin{tabular}{l l c c}        % centered columns (4 columns)
\hline\hline                 % inserts double horizontal lines
Telescope & Instrument & Spectral Range (\AA) & ADC \\    % table heading 
\hline                        % inserts single horizontal line
Keck     & DEIMOS & 4100--11000 & no \\
Keck     & LRIS   & 3100--10000 & planned \\
Subaru   & FOCAS  & 3650--9000  & yes \\
Gemini   & GMOS   & 3600--11000 & yes \\
LBT      & MODS   & 3300--11000 & yes \\
HET      & LRS    & 4150--9100  & yes \\
SALT     & PFIS   & 3200--8500  & yes \\
GTC      & OSIRIS & 3650--10000 & evaluating \\
VLT      & FORS   & 3300--11000 & yes \\
VLT      & VIMOS  & 3300--11000 & no \\
\hline                                   %inserts single line
\end{tabular}
\end{table}

In Table \ref{teltable} I review all current and known future 8m
telescopes and optical MOS spectrographs. There are only 2 
instruments without an ADC in operation or in planning:

DEIMOS on Keck is heavily
red optimized, thus, atmospheric dispersion is not a significant
issue for many projects. The DEIMOS Slitmask design page
(http://www.ucolick.org/$\sim$phillips/deimos\_ref/masks.html) 
provides preliminary tools to evaluate the effect of atmospheric
dispersion on slit loss for short integrations. No guidelines are
provided for long integrations.

The VIMOS manual (\cite{vimos2005}) discusses the effect of
atmospheric dispersion in MOS mode. They arrive at the
conclusion that the only generic way to minimize slit losses is
to orient the slits North-South and observe within $\pm$2 hours
of the meridian. Even though the detailed study of the effect of
atmospheric dispersion on VIMOS (Cuby et al. \cite{cuby1996})
makes no explicit statement about deviating from these
constraints, a casual reading of the manual by an unexperienced
observer may leave the impression that this slit orientation is
the \emph{only} valid strategy.

This conclusion is clearly valid for the sample observation
used in the manual: observing between airmasses of 1.7 and 1.4 in the UV/blue.
On the other hand, for some observing projects, e.g. limiting
UV/blue spectroscopy to low airmass, the advantage of N-S orientation
diminishes. Thus, an additional freedom is available in \emph{some}
cases to maximize scientific return of the observations.

\section{Sample observation}

Throughout this paper, I will use a sample observation of the
Lockman Hole ($\delta$=+57:35:25.0, J2000) using the Keck telescope
(latitude of +19:46:36). I assume multislit spectroscopy
with 1.0" slit width and 1.2" seeing. I assume an ambient
temperature of 2.5 degrees C, ambient pressure of 61.5 kPa (615 mbar) and
a relative humidity of 40\%.

The observations consist of 1 night long integration on a mask,
which for this field implies an hour angle range between
18 hours and 3 hours (the asymmetry is due to the mechanical constraints
of the telescope), which covers an airmass range of 1.27 (hour angle
of 0) to 3.5 (hour angle of 18 hours). I will concentrate on the DEIMOS multiobject
spectrograph, which is red-optimized. Thus, I will concentrate
on the 4500\dots9500 {\AA} wavelength range.

In the calculation I assume that the seeing does \emph{not} depend
on the wavelength and I also assume that the alignment and guiding
is done in the R-band (approximately 7000 {\AA}).

%__________________________________________________________________

\section{Atmospheric dispersion}

The most up-to-date atmospheric dispersion determination, the Ciddor
formula (\cite{ciddor1996}) is reviewed in the Appendix. The most important
formulas are:

The \emph{differential} refraction (as a function of wavelength, relative
to the alignment/guiding effective wavelength, $\lambda_0$) in radians is:
\begin{equation}
\Delta R(\lambda)\equiv R(\lambda)-R(\lambda_0)=(n(\lambda)-n_0)\tan z_a
\end{equation}

\begin{figure}
   \centering
   \includegraphics[width=8cm]{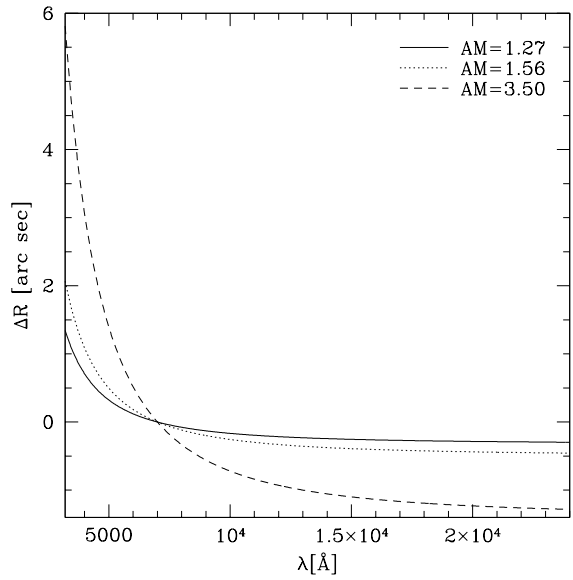}
   \caption{Differential refraction as a function of wavelength
at different airmasses: Solid line -- AM=1.27 (HA=0), dotted line --
AM=1.56 (HA=3h), short dashed line -- AM=3.50 (HA=18h).
I used our sample observation
of the Lockman Hole using Keck (see section \ref{intro} for details).}
\label{delta_r_plot}
    \end{figure}

In figure \ref{delta_r_plot} I plot
$\Delta R$ as a function of wavelength at different wavelengths for
different airmasses.
%                                     One column figure (place early!)
%______________________________________________ Gamma_1 (lg rho, lg e)

The index of refraction, $n_{as}$ of standard air is
\begin{equation}
10^8(n_{as}-1)={5792105\mu m^{-2}\over 238.0185\mu m^{-2}-\sigma^2}+
{167917\mu m^{-2}\over 57.362\mu m^{-2}-\sigma^2}
\end{equation}
where $\sigma$ is the wave number (reciprocal of the
\emph{vacuum} wavelength) in inverse micrometers.

If we are only interested in differential refraction, we can write
a simpler formula that is
sufficiently accurate for many applications:
\begin{equation}
\Delta R(\lambda)\approx {pT_0\over p_0T}(n_{as}(\lambda)-n_{as}(\lambda_0))\tan z_a
\end{equation}

\begin{figure}
\centering
\includegraphics[width=8cm]{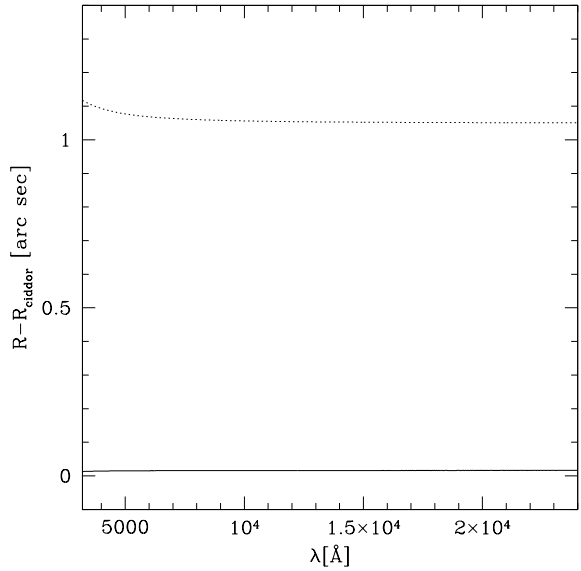}
\caption{
The error introduced by our approximate formula (solid line) and the
old Edl\'en formula (dotted line) in the atmospheric refraction assuming
$\tan z_a=1$.}
\label{checks}
\end{figure}

The error introduced by this approximation (as well as the error introduced
by the old Edl\'en formula) is shown is figure \ref{checks}. As one can see,
even the \emph{absolute} refraction is well reproduced by this simpler formula,
while the old Edl\'en formula is significantly different. For the
calculation of the \emph{differential} refraction, there is no practical
difference between the formulas in the wavelength range considered.

\section{Slit loss}

Now that we have the atmospheric dispersion, we can also calculate 
the slit loss. I assume a point source that has a surface brightness
profile of
\begin{equation}
\mu(r)={1\over2\pi\sigma^2}e^{-r^2/2\sigma^2}
\end{equation}

If we assume a slit of width $2a$, that is sufficiently long,
the fraction of light entering the slit from an object that is
displaced by $x_0$ \emph{perpendicular} to the slit (a displacement
parallel to the slit does not affect the amount of light entering
the slit) is
\begin{equation}
I(x_0)={1\over\sigma\sqrt{2\pi}}\int\limits_{-a+x_0}^{a+x_0} e^{-x^2/2\sigma^2}dx
\end{equation}

\begin{figure*}
\centering
\includegraphics[width=8cm]{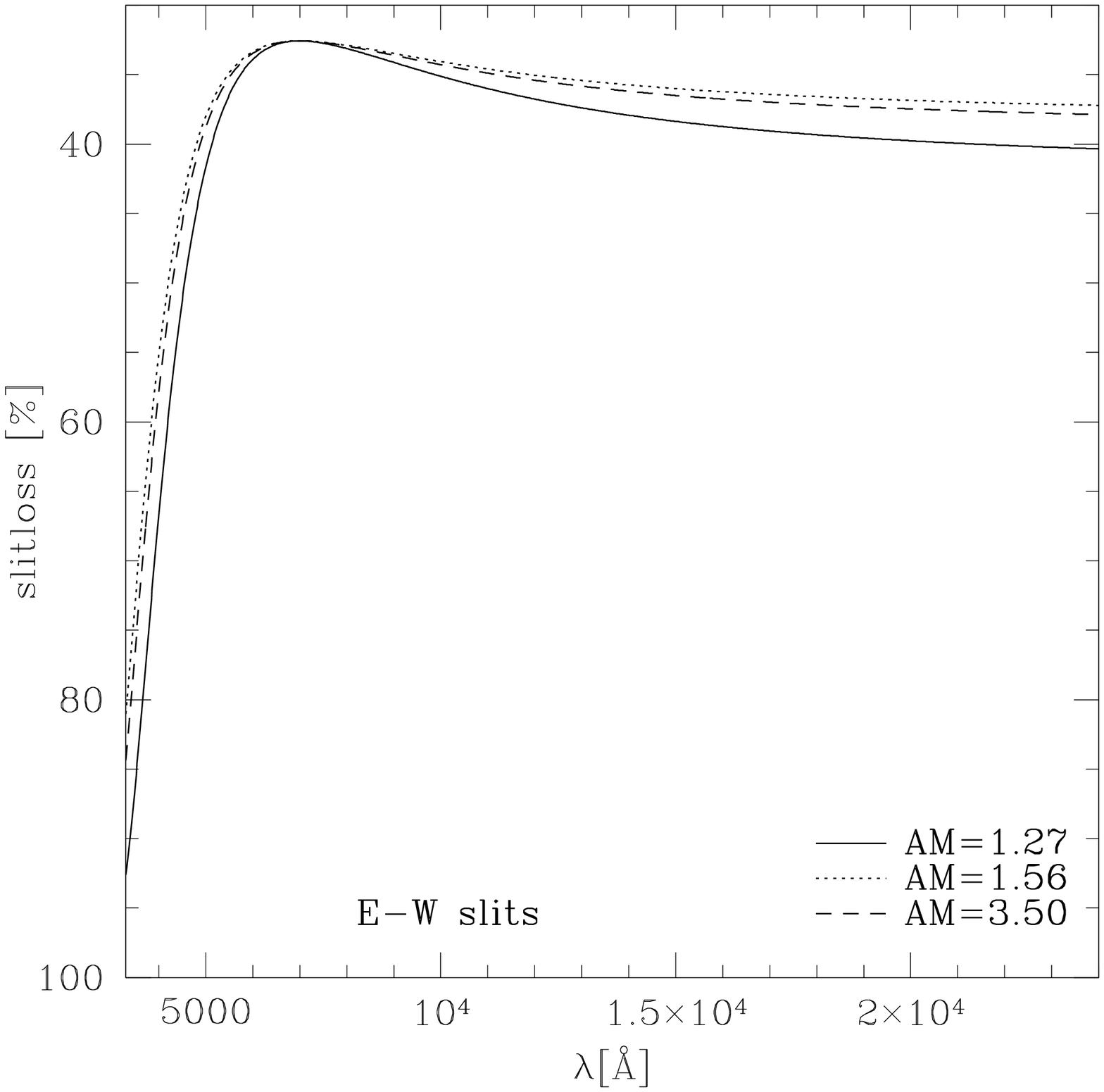}
\includegraphics[width=8cm]{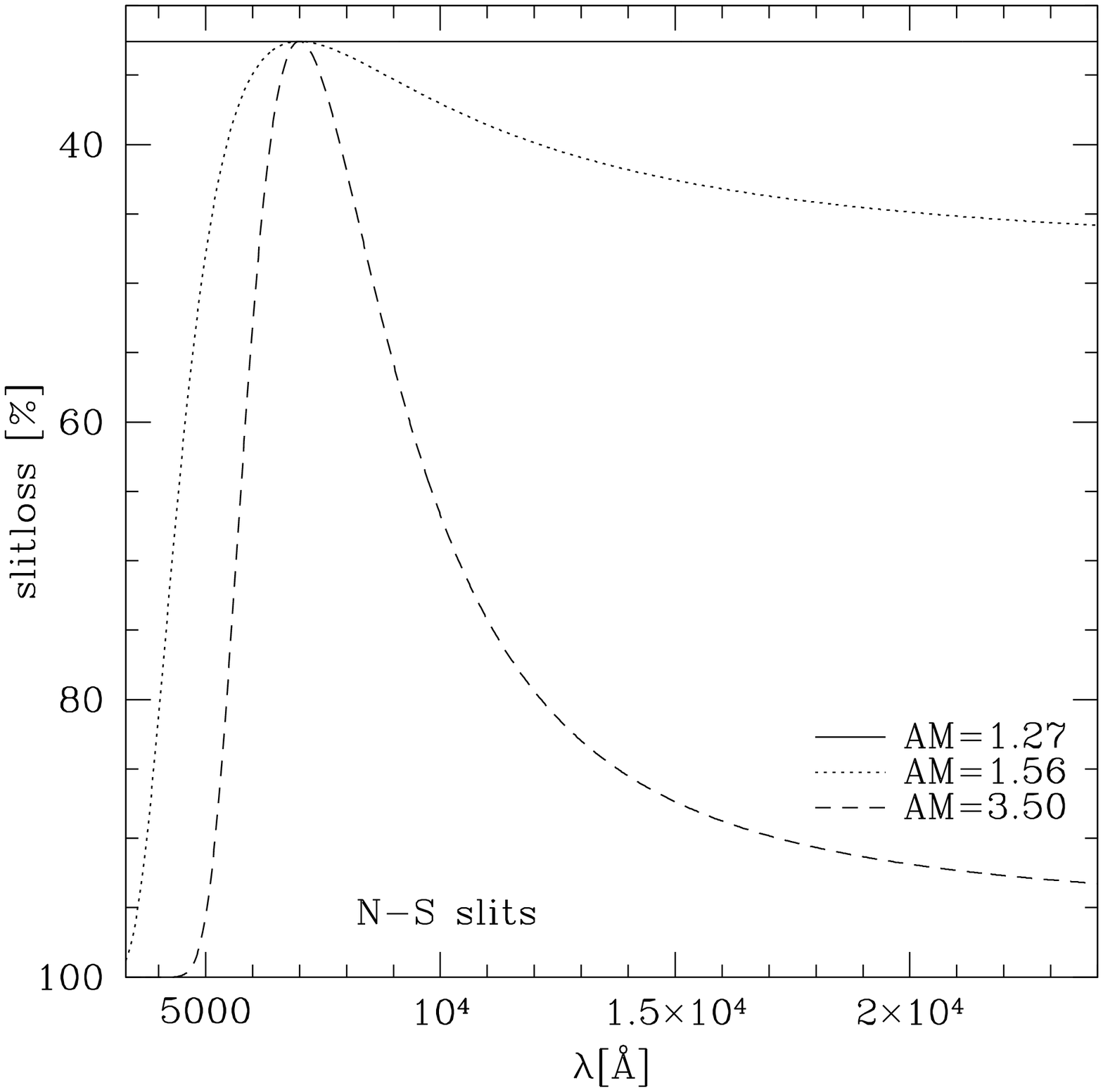}
\caption{Slit loss as a function of wavelength
at different airmasses. I used our sample observation
of the Lockman Hole using Keck (see section \ref{intro} for details).
Notice the different scales used on the plots.}
\label{slitloss}
\end{figure*}

The perpendicular displacement, $x_0$, depends on the differential
refraction, $\Delta R$, and the angle between the slit and and
parallactic angle. In figure \ref{slitloss} I consider two
configurations, an East-West oriented slit (i.e. the slits are
perpendicular to the parallactic angle at the meridian) and a North-South
orientation (slits are parallel to the parallactic angle at the meridian).

As we can see, the N-S slit orientation results in a low slit loss
at hour angle of 0 (minimal airmass) that does not depend on
wavelength (as we are observing close to the parallactic angle).
On the other hand, as we are moving away from the optimal
configuration, the situation deteriorates rapidly. This is due to the
fact that the atmospheric dispersion increases as the airmass
increases \emph{and} the slit orientation is moving away from
the ideal, parallactic angle -- both effects increase the slit loss.

In the alternative configuration, i.e. East-West slit orientation,
the slit loss is never optimal. Even at low airmass, a significant
fraction of the light is lost (e.g. slit loss is 42\% at 5000 {\AA},
instead of 33\%), \emph{but} the slit loss does not deteriorate
so quickly. This is due to the fact that as the airmass increases,
the dispersion increases, but the slit is getting closer to
the parallactic angle, thus the \emph{projected} dispersion
is not increasing so rapidly. In fact, in our particular configuration
(Lockman Hole and Keck), the slit loss is actually smaller at
high airmass (42\%, 38\% and 39\% at 5000 {\AA} at the airmass of
1.27, 1.56 and 3.50, respectively).

\begin{figure}
\centering
\includegraphics[width=5cm]{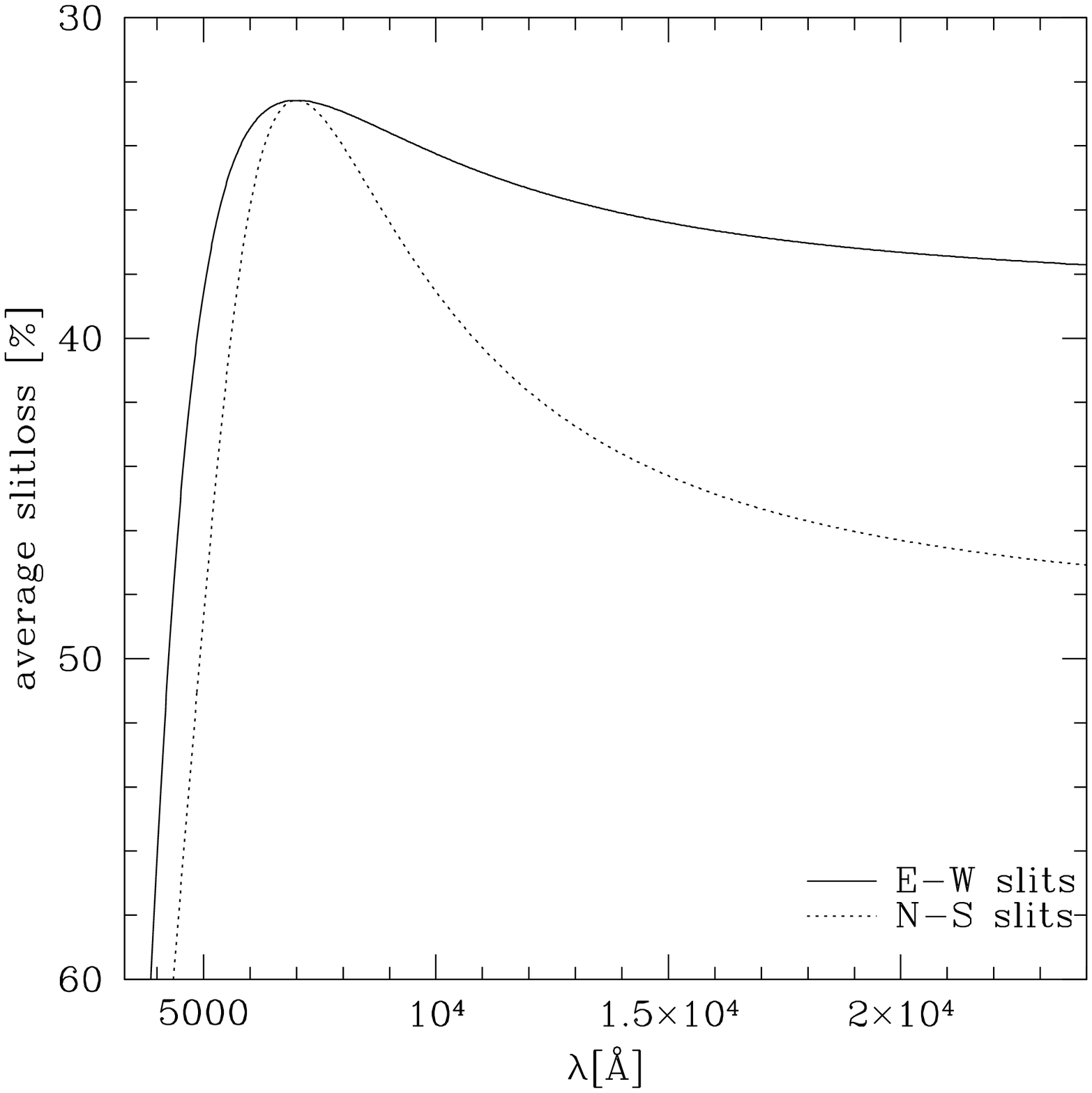}
\caption{Average slit loss at different slit orientations. Solid line
indicates East-West slit orientation, dotted lines shows North-South
orientation.  I used our sample observation
of the Lockman Hole using Keck (see section \ref{intro} for details).}
\label{average}
\end{figure}

To evaluate the overall effect of the slit orientation on the
signal level achievable, we can also calculate the `average'
slit loss of a long exposure. This is shown in figure \ref{average}.

%
%______________________________________________________________

\begin{figure}
\centering
\includegraphics[width=15cm]{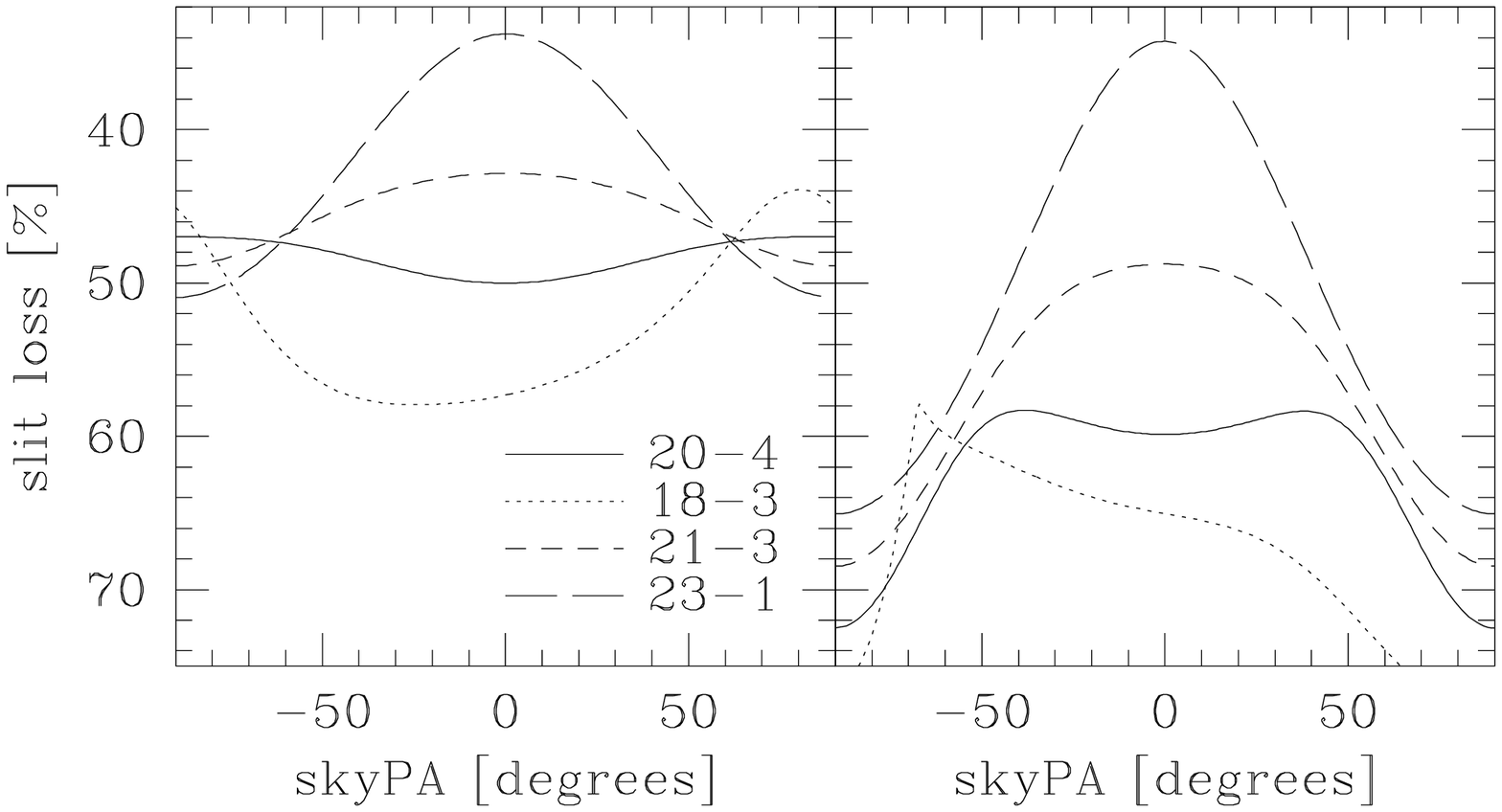}
\caption{Average slit loss at 4500{\AA} at different slit orientations. 
I used our sample observation
of the Lockman Hole using Keck (see section \ref{intro} for details -- left panel) and a southern field (CDFS, -27:48:30 declination -- right panel).
I show the effect of slit sky position angle at different hour angle
ranges. For sky position angle I use the (unusual) DEIMOS definition,
where 90 degree is E-W, 0 degree is N-S.}
\label{pa}
\end{figure}

In figure \ref{pa} I show the effect of sky position angle
on average slit loss at 4500{\AA}, using my example observations
and a southern field (Chandra Deep Field South).
As we can see, for short exposures around meridian passage,
the N-S orientation is optimal. As we go to longer and longer
integration times, the effect of slit orientation becomes smaller
and smaller, while at extremely long integration times, the
E-W orientation becomes ideal for the Lockman Hole field. For asymmetric
cases (for example
our actual observations between hour angles 18 and 3) the optimal
slit orientation is neither E-W nor N-S.

In contrast, for the southern field, the optimal slit orientation
remains North-South. This is due to the fact that for the CDFS
field the airmass is never below 1.48 from Keck, thus, the
differential refraction is comparable to the slit width even
at meridian passing.

%
%______________________________________________________________

\section{Field differential refraction\label{fieldref}}

All results presented so far only hold for objects in the center of
the field of view of the instrument. 
For wide-field spectrographs, an additional effect is important:
the zenith distance is not constant across the field of view, thus
the differential refraction is not constant, either. This is
an achromatic effect, which we can calculate using equation \ref{R1}:

\begin{equation}
\delta R\approx (n-1) {d \tan z_a\over dz_a}\delta z_a=
(n-1)\sec^2 z_a\delta z_a
\end{equation}
where $\delta R$ is the variation in the differential refraction
across the field and $\delta z_a$ is the field of view. As
we have seen, $(n-1)$ is on the order of $3\times10^{-4}$, 
$\sec z_a$ is on the order of 1, thus a 1000 arc second field
of view (nearly 17 arc minutes) introduces a \emph{field} differential
refraction on the order of 0.3 arc second. It is important to point
out that this effect makes it imperative to realign the masks
periodically unless guiding is near the center of the field or
the telescope compensates for off-axis guiding.

\section{Web interface}

I make our code used in our calculations available to the
community in both source code and web application form at
{\bf http://www.xray.mpe.mpg.de/$\sim$szgyula/slitloss/}. The program uses
the full Ciddor formula
to evaluate the slit loss, but for comparison all three
formulas are available to calculate the differential refraction.

%
%______________________________________________________________

\section{Conclusions}

I demonstrated that choosing the optimal slit orientation for
multiobject spectroscopy, using long exposure times, requires
care and should be evaluated individually for each project. For
every field and expected duration, one has to find a balance.

I reviewed the most recent determination of atmospheric dispersion.
For typical cases, I found that a simplified version of the
most up-to-date Ciddor formula can be
used, due to the fact that alignment/guiding removes the effect
of dispersion in zeroth order. The simplified formula only depends
on pressure and temperature. The effect of relative humidity and
CO$_2$ concentration is very small. Furthermore, the differential
refraction follows very simple scaling rules, i.e. it scales linearly
with pressure and the inverse of temperature (in Kelvins).

For short exposures, the optimal strategy is, as expected, still
to orient the slits with the parallactic angle. On the other hand, 
for longer exposures, this is not always the right strategy. There
are two effects to consider, the increasing differential refraction
and the changing angle between the slits and the parallactic angle.
Depending on the configuration, these effects can work against
each other, thus resulting in a long, relatively stable observation
that is never optimal or these can amplify each other, thus resulting
in an optimal short observation that deteriorate very fast.

It is also important to point out that alignment/guiding is crucial.
One has to select the effective wavelength of these to maximize the
science output. It is absolutely worthwhile to spend a few extra minutes
every few hours using some standard filters instead of using
no filters at all, especially with alignment stars with unknown
spectral types. This latter approach runs the risk of using
very blue stars for alignment, thus the alignment will only be
optimal in the blue, where one may not be observing.

Naturally, in the long run, the use of atmospheric dispersion
correctors should be considered. As I have shown, these can
improve the throughput by as much as a factor of two for instruments
operating in the blue. The
cost of these from an observational point of view is small (few photons are
lost in the extra optical elements) as
only two very weak prisms are sufficient in most cases to
produce a `tunable' prism to compensate for differential refraction
in first order. This can solve the problem of differential refraction,
but does not eliminate the slit loss completely. Finite slit widths
will always `cut' the object signal. As the seeing can be wavelength
dependent (this effect was completely ignored in this paper), 
so can the slit loss be wavelength dependent. Thus, accurate
spectrophotometry still requires very wide slits, and consequently,
very low spectral resolution and a significantly degraded signal to
noise ratio.

Finally, I provide a web based service to the community and we
also release the software developed.

\begin{acknowledgements}
      Part of this work was supported by the German
      \emph{Deut\-sche For\-schungs\-ge\-mein\-schaft, DFG\/} project
      number Ts~17/2--1.
\end{acknowledgements}

\appendix

\section{Atmospheric dispersion\label{appendix}}

Atmospheric dispersion (i.e. the apparent displacement of object), $R$,
is defined as
\begin{equation}
R=z_t-z_a
\end{equation}
where $z_t$ and $z_a$ are the true and apparent zenith distances,
respectively.

Assuming that the index of refraction depends only on height, using
Snell's law, we can write that 
\begin{equation}
n(h)\sin(z(h))\equiv const.
\end{equation}
where $h$ is the height, $z(h)$ is the apparent zenith distance at
height $h$.
\begin{equation}
n \sin z_a=\sin z_t
\end{equation}
Thus, the apparent zenith distance at the telescope
only depends on the index of refraction at the observatory. Assuming
that $R$ is small, i.e. $\sin R\approx R$ and $\cos R\approx 1$,
we can write
\begin{equation}
R\approx (n-1)\tan z_a\label{R1}
\end{equation}

As no telescope points accurately enough (i.e. with less than
0.1 arc second accuracy required by slit based spectroscopy), all
observations start with an `alignment'. This step guarantees that
all objects are centered on the slit \emph{at a particular wavelength},
$\lambda_0$ (determined by the filter used for the alignment).
During the exposure, the guiding subsystem and periodic realignments
maintain this condition. As the guider typically does not operate at
the same wavelength, special care is required to compensate for
atmospheric refraction in the guider system. Even with an ideal
guider, the mask gets misaligned due to other effects, e.g. the
open loop instrument rotator (especially on modern, altitude-azimuth
mounted telescopes). Thus, a periodic realignment is mandatory.
For simplicity, I will assume that the telescope is equiped with
an ideal alignment and guiding system, thus the objects are always
centered on the slit at wavelength $\lambda_0$. In section \ref{fieldref}
I discuss why this can not hold for wide field of view instruments.

As a consequence, the \emph{absolute} magnitude of
atmospheric refraction is irrelevant, since alignment/guiding
automatically compensates for it. The relevant quantity, the
\emph{differential} refraction (as a function of wavelength, relative
to the alignment/guiding effective wavelength, $\lambda_0$) in radians is:
\begin{equation}
\Delta R(\lambda)\equiv R(\lambda)-R(\lambda_0)=(n(\lambda)-n_0)\tan z_a
\end{equation}

The calculation of the refractive index is a crucial part of our
calculation. Unfortunately, there are still old formulas in use,
most notably the Cauchy formula and the old and new Edl\'en formulas
that are at least 50 years old. These formulas are known to be
inaccurate, but they still crop up in the literature and astronomical
applications. The current best formula is the Ciddor (\cite{ciddor1996})
formula, presented below.

The index of refraction, $n_{as}$ of standard air, i.e. dry air at 15$^\circ$C
temperature, using the International Temperature Scale of 1990 (Saunders, \cite{saunders1990})
, 101325 Pa pressure and 450 ppm (part per million)
CO$_2$ concentration, is
\begin{equation}
10^8(n_{as}-1)={5792105\mu m^{-2}\over 238.0185\mu m^{-2}-\sigma^2}+
{167917\mu m^{-2}\over 57.362\mu m^{-2}-\sigma^2}
\end{equation}
where $\sigma$ is the wave number (reciprocal of the
\emph{vacuum} wavelength) in inverse micrometers. In the range
of 3500\AA\dots 24000\AA, $10^8(n_{as}-1)$ is in the range of
28612\dots27289.

If the CO$_2$ concentration is $x_c$ ppm instead of 450 ppm, the
index of refraction, $n_{axs}$, is
\begin{equation}
n_{axs}-1=(n_{as}-1)\left(1+0.534\times10^{-6}\left(x_c-450\right)\right).
\end{equation}
This formula is accurate to $10^{-8}$ for the refractive index up to
600 ppm CO$_2$ concentrations 
in the range of 360-2500 nm. In this range,
the effect of CO$_2$ variation on $n_{axs}$ is on the order of $10^{-7}$.

For water vapor at the `standard conditions', i.e. at 20 $^\circ$C and
1333 Pa, the index of refraction, $n_{ws}$, is
\begin{eqnarray}
10^8(n_{ws}-1)=1.022(295.235\mu m^{-2}+2.6422\mu m^{-2}\sigma^2 - \nonumber\\
0.032380\mu m^{-4}\sigma^4+0.004028\mu m^{-6}\sigma^6)
\end{eqnarray}
The formula is accurate to $2\times 10^{-7}$ in the range of 350-1200 nm.
In the optical/near-IR range $\sigma$ is between 0.5 and 3. Thus, the
value of $10^8(n_{ws}-1)$ is 324\dots302 in this wavelength range.

The saturation vapor pressure of water vapor, $p_{vs}$, at temperature $T$
(in Kelvins), over liquid water is
\begin{eqnarray}
p_{vs} = \exp(1.2378847\times 10^{-5} K^{-2} T^2 -\nonumber \\
1.9121316\times10^{-2} K^{-1} T +33.93711047-6343.1645K/T)
\end{eqnarray}
Considering a temperature range of -20 $^\circ$C\dots 40 $^\circ$C,
the saturated vapor pressure is 0.1\dots7.4 kPa.

The enhancement factor of water vapor in air is
\begin{equation}
f=1.00062+3.14\times10^{-8} Pa^{-1} p+ 5.6\times10^{-7}\mbox{}^\circ C^{-2} t^2
\end{equation}
where $p$ is the pressure and $t=T-273.15K$. The deviation of $f$ from 1
is at most $4\times10^{-4}$.

The molar fraction of water vapor in moist air is
\begin{equation}
x_w={f\,h\,p_{vs}\over p}
\end{equation}
where $h$ is the fractional humidity (between 0 and 1). The range of
$x_w$ is 0\dots0.25, but the high value (0.25) is assuming unrealistic
conditions, i.e. 40 $^\circ$C temperature, 100\% humidity and
Mt. Everest type ambient pressure (30 kPa). In most cases, accepting
a range of 0\dots0.05 (20 $^\circ$C, 60 kPa) is more realistic.

The compressibility of the moist air, $Z$ is
\begin{eqnarray}
\lefteqn{Z=}\nonumber\\
&&1-{p\over T}[1.58123\times10^{-6} K\,Pa^{-1}-2.9331\times10^{-8}Pa^{-1}t+ \nonumber\\
&& 1.1043\times10^{-10}K^{-1}\,Pa^{-1}t^2+\nonumber\\
&& \left(5.707\times10^{-6}K\,Pa^{-1}-2.051\times10^{-8}Pa^{-1}t\right)x_w+\nonumber\\
&& \left(1.9898\times10^{-4}K\,Pa^{-1}-2.376\times10^{-6}Pa^{-1}t\right)x_w^2]+\nonumber\\
&& \left({p\over t}\right)^2\left(1.83\times10^{-11} K^2 Pa^{-2}-0.765\times10^{-8}K^2Pa^{-2}x_w^2\right)
\end{eqnarray}
where $p$ is pressure (in Pascals), T is temperature (in Kelvins)
and $t=T-273.15K$. Considering realistic pressures (30\dots100 kPa),
temperatures (-20\dots 40 $^\circ$C) and water vapor molar fractions
($x_w<0.1$), the compressibility is very close to one: $\vert Z-1\vert<
2\times10^{-3}$ (using a very conservative upper limit estimate).

At standard conditions ($p=101325Pa$, t=15\mbox{}$^\circ$ C, dry air), the
standard compressibility is
\begin{equation}
Z_0\approx 0.9995922115
\end{equation}

At the saturated water vapor conditions, i.e. $p=1333 Pa$ and $t=20\mbox{}^\circ C$, the compressibility is
\begin{equation}
Z_1\approx 0.9999952769
\end{equation}

The molar mass of air is
\begin{equation}
M_a=10^{-3}\left(28.9635+12.011\times10^{-6}\left(x_c-400\right)\right)
\end{equation}
in kg/mol units ($x_c$ is the CO$_2$ concentration in ppm, as used above).

The density of air, $\rho$ (in kg/m$^3$ units) is
\begin{equation}
\rho={pM_a\over ZRT}\left(1-x_w\left(1-{M_w\over M_a}\right)\right)
\end{equation}
where $R=8.314510 J\,mol^{-1}\, K^{-1}$, the gas constant, $M_w=0.018015kg/mol$,
the molar mass of water vapor.

At standard conditions ($p_0=101325 Pa$ and $t_0=15\mbox{}^\circ$C) the 
density of dry air ($x_w=0$) only depends on the CO$_2$ concentration
\begin{equation}
\rho_{axs}={p_0M_a\over Z_0RT_0}={28.9635\times 10^{-3}3p_0\over Z_0RT_0}
\left(1+{12.011\over28.9635}\times{x_c-400\over10^6}\right)
\end{equation}

At water vapor standard conditions ($p_1=$1333 Pa, $t_1=$20 $^\circ$C, $w_s=1$), the
saturated water vapor density is
\begin{equation}
\rho_{ws}={p_1 M_w\over Z_1 R T_1}\approx 0.00985235
\end{equation}

For actual conditions, the air density of the air component is
\begin{equation}
\rho_a={pM_a\over ZRT}\left(1-x_w\right)
\end{equation}
and
\begin{equation}
{\rho_a\over\rho_{axs}}={Z_0\over Z} {p\over p_0} {T_0\over T}(1-x_w)
\end{equation}

The water vapor component is
\begin{equation}
\rho_v={pM_wx_w\over ZRT}
\end{equation}
and
\begin{equation}
{\rho_v\over\rho_{ws}}={p\over p_1} {T_1\over T} {Z_1\over Z}x_w=
{T_1\over T} {Z_1\over Z}{p_{vs}\over p_1} f h
\end{equation}
Considering that $Z$ and $f$ are very close to one, assuming realistic
ranges for the parameters, we can place a very conservative upper
limit on the vapor density to standard saturated water vapor density
ratio: $\rho_v/\rho_{ws}<$6.5.

Finally, the refractive index is
\begin{equation}
n-1={\rho_a\over\rho_{axs}}\left(n_{axs}-1\right)+{\rho_v\over\rho_{ws}}\left(n_{ws}-1\right)
\end{equation}

As I have shown above, $n_{ws}$ varies by at most $2\times10^{-7}$ as
a function of wavelength in the optical/near-IR range and 
$\rho_v/\rho_{ws}<$6.5. Thus, the second,
water related term can change by at most $1.3\times10^{-6}$, thus the
differential refraction can change by at most this much. This 
limits the effect of water in the atmosphere to 0.3 arc seconds in 
the most extreme case: alignment/guiding in the K-band, observing in the UV,
close to 100\% humidity. In a realistic case, the effect is much
smaller so the second term can be ignored most of the time. The first term
in the equation \emph{does depend} on humidity through the air density,
but in realistic cases the only water related term, $1-x_w$ varies
by a few percent, thus the effect is very small.

As I have shown, the variation in $n_{axs}$ introduced by CO$_2$
concentration variation is less than $10^{-7}$. Thus, the 
differential refraction variation introduced is 0.03 arc seconds
or less. For most applications, this is negligible.

If we are only interested in differential refraction and are considering
these approximations, we can write a much simpler formula that is
sufficiently accurate for many applications:
\begin{equation}
\Delta R(\lambda)\approx {pT_0\over p_0T}(n_{as}(\lambda)-n_{as}(\lambda_0))\tan z_a
\end{equation}

In Table \ref{modeltable} the importance of atmospheric condition variations 
is shown. The effect of relative humidity and CO$_2$ concentration is
negligible in most observations.

\begin{table}
\caption{The effect of different environmental parameters on the differential
refraction. I show the effects of changes in the atmospheric parameters
on the differential refraction using. I use our observations of the
lockman hole with Keck. As a reference point, I assume 2.5 $^\circ$C
temperature, 61.5 kPa ambient pressure, 40\% relative humidity
and 450 ppm CO$_2$ concentration. The effect of changing \emph{one} of
these four parameters (while keeping the other three at the nominal
value) are shown below. Only the differential refraction for 3500{\AA}
is shown, $\Delta R(3500\hbox{\AA})$ in arc seconds. The hour angle is
18h.}
\label{modeltable}      % is used to refer this table in the text
\centering                          % used for centering table
\begin{tabular}{l c}        % centered columns (4 columns)
\hline\hline                 % inserts double horizontal lines
Parameter change & $\Delta R(3500\hbox{\AA})$ \\    % table heading 
\hline                        % inserts single horizontal line
nominal & 4.53\\
t=-10 $^\circ$C & 4.75 \\
t=20 $^\circ$C & 4.27 \\
p=30kPa & 2.21 \\
p=100 kPa & 7.37 \\
RH=0\% & 4.53 \\
RH=90\% & 4.54 \\
$x_c$=300 ppm & 4.53 \\
$x_c$=500 ppm & 4.53 \\
\hline                                   %inserts single line
\end{tabular}
\end{table}

\end{document}